\documentclass{natureJLH}
\usepackage{graphicx}
\usepackage[none]{hyphenat}
\usepackage{tikz}
\usepackage{ulem} 
\usepackage{lineno}
\usepackage{amssymb}
\usepackage{amsmath}
\usepackage{caption}
\usepackage{booktabs,multirow}

\title{A wide star-black-hole binary system from radial-velocity measurements}

\author{Jifeng Liu$^{\star1,2,3}$, Haotong Zhang$^{\star1}$, Andrew W. Howard$^4$, Zhongrui Bai$^1$, Youjun Lu$^{1,2}$, Roberto Soria$^{2,5}$, Stephen Justham$^{2,1,6}$, Xiangdong Li$^{7,8}$, Zheng Zheng$^{9}$, Tinggui Wang$^{10}$, Krzysztof Belczynski$^{11}$, Jorge Casares$^{12,13}$, Wei Zhang$^1$, Hailong Yuan$^1$, Yiqiao Dong$^1$, Yajuan Lei$^1$, Howard Isaacson$^{14}$, Song Wang$^1$, Yu Bai$^1$, Yong Shao$^{7,8}$, Qing Gao$^1$, Yilun Wang$^{1,2}$, Zexi Niu$^{1,2}$, Kaiming Cui$^{1,2}$, Chuanjie Zheng$^{1,2}$, Xiaoyong Mu$^2$, Lan Zhang$^1$, Wei Wang$^{15,3}$, Alexander Heger$^{16}$, 
Zhaoxiang Qi$^{17,1}$, Shilong Liao$^{17}$, Mario Lattanzi$^{18}$
Wei-Min Gu$^{19}$, Junfeng Wang$^{19}$, Jianfeng Wu$^{19}$, Lijing Shao$^{20}$, Rongfeng Shen$^{21}$, Xiaofeng Wang$^{22}$, Joel Bregman$^{23}$, Rosanne Di Stefano$^{24}$, Qingzhong Liu$^{25}$, Zhanwen Han$^{26}$, Tianmeng Zhang$^1$, Huijuan Wang$^1$, Juanjuan Ren$^1$, Junbo Zhang$^1$, Jujia Zhang$^{26}$, Xiaoli Wang$^{26}$, Antonio Cabrera-Lavers$^{12,27}$, Romano Corradi$^{12,27}$, Rafael Rebolo$^{13,27}$, Yongheng Zhao$^{1,2}$, Gang Zhao$^{1,2}$, Yaoquan Chu$^{10}$, and Xiangqun Cui$^{28}$. }

\begin{document}

\maketitle

\begin{affiliations}

\item Key Laboratory of Optical Astronomy, National Astronomical Observatories, Chinese Academy of Sciences, Beijing, China
\item School of Astronomy and Space Sciences, University of Chinese Academy of Sciences, Beijing, China
\item WHU-NAOC Joint Center for Astronomy, Wuhan University, Wuhan, China
\item Department of Astronomy, Caltech, Pasadena, CA, USA
\item Sydney Institute for Astronomy, School of Physics A28, The University of Sydney, Sydney, New South Wales, Australia
\item The Anton Pannekoek Institute for Astronomy, University of Amsterdam, Amsterdam, The Netherlands
\item School of Astronomy and Space Science, Nanjing University, Nanjing, China
\item Key Laboratory of Modern Astronomy and Astrophysics (Nanjing University), Misnistry of Education, Nanjing, China
\item Department of Physics and Astronomy, University of Utah, Salt Lake City, UT, USA
\item CAS Key Laboratory for Research in Galaxies and Cosmology, Department of Astronomy, University of Science and Technology of China, Hefei, China
\item Nicolaus Copernicus Astronomical Centre, Polish Academy of Sciences, Warsaw, Poland
\item Instituto de Astrof\'{i}sica de Canarias, La Laguna, Tenerife, Spain
\item Departamento de Astrofísica, Universidad de La Laguna, Santa Cruz de Tenerife, Spain
\item Astronomy Department, University of California, Berkeley, CA, USA
\item School of Physics and Technology,  Wuhan University, Wuhan, China
\item Monash Centre for Astrophysics, School of Physics and Astronomy, 19 Rainforest Walk, Monash University, Victoria, Australia
\item Shanghai Astronomical Observatory, Chinese Academy of Sciences, Shanghai, China
\item INAF--Osservatorio Astrofisico di Torino, Pino Torinese, Italy
\item Department of Astronomy, Xiamen University, Xiamen, China
\item Kavli Institute for Astronomy and Astrophysics, Peking University, Beijing, China
\item School of Physics and Astronomy, Sun Yat-Sen University, Zhuhai, China
\item Physics Department and Tsinghua Center for Astrophysics, Tsinghua University, Beijing, China
\item Department of Astronomy, University of Michigan, 1085 South University Street, Ann Arbor, MI, USA.
\item Harvard-Smithsonian Center for Astrophysics, Cambridge, MA, USA
\item Key Laboratory of Dark Matter and Space Astronomy, Purple Mountain Observatory, Chinese Academy of Sciences, Nanjing, China
\item Key Laboratory for the Structure and Evolution of Celestial Objects, Yunnan Observatories, Chinese Academy of Sciences, Kunming, China
\item GRANTECAN, Cuesta de San Jos\'{e} s/n, Bre\~{n}a Baja, La Palma, Spain
\item Nanjing Institute of Astronomical Optics and Technology, Chinese Academy of Sciences, Nanjing, China

\end{affiliations}

\begin{abstract}

All stellar mass black holes have hitherto been identified by X-rays emitted by gas that is accreting onto the black hole from a companion star. These systems are all binaries with black holes below $30$ M$_\odot$\cite{2006csxs.book..157M,2014Natur.505..378C,2008ApJ...678L..17S,2010MNRAS.403L..41C}. Theory predicts, however, that X-ray emitting systems form a minority of the total population of star-black hole binaries\cite{Romani1998,2009ApJ...707..870B}. When the black hole is not accreting gas, it can be found through radial velocity measurements of the motion of the companion star. Here we report radial velocity measurements of a Galactic star, LB-1, which is a B-type star, taken over two years. We find that the motion of the B-star and an accompanying H$\alpha$ emission line require the presence of a dark companion with a mass of $68^{+11}_{-13}$ M$_\odot$, which can only be a black hole. The long orbital period of 78.9 days shows that this is a wide binary system. The gravitational wave experiments have detected similarly massive black holes\cite{Abbott2016,2018arXiv181112907T}, but forming such massive ones in a high-metallicity environment would be extremely challenging to current stellar evolution theories\cite{2003ApJ...591..288H, Belczynski2010b, 2015MNRAS.451.4086S}.

\end{abstract}

A radial-velocity monitoring campaign with the Large Aperture Multi-Object Spectroscopic Telescope\cite{2012RAA....12.1197C} (hereafter LAMOST) has been carried out to discover and study spectroscopic binaries since 2016, and has obtained 26 measurements each for about 3000 targets brighter than 14 mag in the Kepler K2-0 field of the sky\cite{2014PASP..126..398H}. One of the B-type stars toward the Galactic Anti-Center, hereafter LB-1, located at $(l,b)=(188.23526,+02.05089)$ with V magnitude of $\sim$11.5 mag, exhibited periodic radial-velocity variation, along with a strong, broad $H_\alpha$ emission line that is almost stationary. Subsequent GTC/OSIRIS\cite{Cepa2000} and Keck/HIRES\cite{Vogt1994} observations between 2017 December and 2018 April have confirmed the periodic variations and the prominent $H_\alpha$ emission line with higher spectral resolution. The spectra reveal three types of lines: stellar absorption lines with apparent periodic motion, a broad $H_\alpha$ emission line moving in anti-phase with much smaller amplitude, and interstellar absorption lines that are time-independent (see Fig.~1).

The overall spectral shape of LB-1 suggests a B-type star characterized by prominent Balmer absorption lines without a significant Balmer jump. The metallicity, as measured from the SiII/MgII lines, is about $1.2\pm0.2$\,Z$_\odot$ ($Z_\odot=0.017$), consistent with that expected for a young B-type star in the Galactic plane. TLUSTY\cite{Hubeny1995} model fitting to the high-resolution Keck spectra leads to $T_{\rm eff} = 18,100\pm820$\,K and ${\rm log}g = 3.43\pm 0.15$, where $g$ is the surface gravity. (The $H_\alpha$ and $H_\beta$ lines were excluded from the fit because of contamination from emission.) Such $T_{\rm eff}$ and ${\rm log}g$ values fit stellar models\cite{Bressan2012} around the main-sequence turn-off points with mass $M_{\rm B} = 8.2^{+0.9}_{-1.2}\,$M$_\odot$, radius $R_{\rm B} = 9\pm2$\,R$_\odot$, and age $t = 35^{+13}_{-7}$\,Myr. The best-fit model is a subgiant B-type star about 0.2\,Myrs after the main-sequence turn-off point.
Its distance $D$ and extinction E(B-V) can be derived simultaneously from fitting its wide-band spectral energy distribution, resulting in $D = 4.23 \pm 0.24$\,kpc and E(B-V) = $0.55\pm0.03$\,mag (see Methods). These values are consistent with the 3D extinction map\cite{Green2015} along LB-1's direction,  so supporting this model. A subdwarf star with a similar temperature is strongly ruled out by the narrow Balmer lines, as shown in Figure~1a, and also by the spectral energy distribution fitting.

The radial motion of the star, as measured from the stellar absorption lines in 26 LAMOST, 21 GTC and 7 Keck observations obtained over two years, can be best fit with a period of $P=78.9\pm0.3$\,days (see Methods). Fitting a binary orbit to the folded radial-velocity curve (see Fig.~2) yields a semi-amplitude $K_{\rm B} = 52.8\pm0.7$\,km/s, an eccentricity $e = 0.03\pm0.01$, and a center-of-mass velocity $V_{\rm 0B} = 28.7\pm0.5$\,km/s. For this binary with a nearly circular orbit, the mass function is 
$PK_{\rm B}^3/2\pi G  = 1.20\pm0.05$\,M$_\odot$, which is the absolute lower limit for the mass of the dark companion to the B star. Given that $M_{\rm B}$ is already known, the minimum mass of the dark primary can be computed as $6.3^{+0.4}_{-1.0}$\,M$_\odot$ for the edge-on geometry with $i = 90^\circ$.
It must be a black hole (BH), since a 6 M$_\odot$ main-sequence star is only about 4--6 times fainter than the B star, and the line features would be easily detected from the Keck spectra.
The BH mass will be 7.8/20/84/245\,M$_\odot$ for lower inclinations at $i=60^\circ$/30$^\circ$/15$^\circ$/10$^\circ$, respectively. The binary separation is about 0.9--2.3\,AU for a BH mass of 6--250\,M$_\odot$, making it a BH binary wider than any previously-known Galatic BH binaries\cite{2006csxs.book..157M,2014Natur.505..378C}.

The prominent $H_\alpha$ emission line is too broad, with a full width at half maximum of 240\,km/s, to arise from an interloper M dwarf or surrounding nebulae, and nor can it be associated with a background AGN/QSO, because this would have other prominent lines at non-zero redshift. Its complicated multi-peak profile (see Fig.~2) suggests an origin from a gaseous Keplerian disk, which can be around the B star, the BH, or the binary. However, the inferred gaseous disk cannot be around the B star, because the $H_\alpha$ emission line is not tracing the motion of the B star, as clearly shown in Fig.~1b. The line profile is distinctly different from a simple double-horned profile for a Keplerian disk viewed at high inclinations. It shows a wine-bottle shape with multiple peaks in the line center, which correspond to substantial non-coherent scattering components from a disk viewed at low inclinations\cite{1996A&AS..116..309H,Hummel1994}. A circumbinary disk would have an inner radius truncated at 1.7 times of the binary separation\cite{Artymowicz1994}, and its corresponding projected velocity is ${1 \over \sqrt{1.7}} \approx 0.75$ times that of the visible star, i.e., about 40\,km/s. The emission line from such a circumbinary disk will be confined to within $\pm40$\,km/s, yet the observed line is three times wider with wings extended beyond $\pm300$\,km/s. This supports that the $H_\alpha$ emission line does not come from a circumbinary disk, but from a disk around the BH.

The BH mass can be obtained directly using the $H_\alpha$ emission line to trace the motion of the BH, and comparing it to the motion of the visible star. The radial velocities of the $H_\alpha$ line, after folding with the period of 78.9\,days, can be fitted with a sinusoid in anti-phase with the visible star. However, the line center may contain contributions from circumbinary materials and accretion spots that are not symmetrically centered on the BH, which would decrease the inferred BH motion and should be masked out. We experimented with different masking schemes, and found that unmasked line wings below 1/3 height can effectively avoid contamination from the line center, yielding a semi-amplitude $K_\alpha = 6.4\pm0.8$\,km/s and a center-of-mass velocity $V_{0\alpha} = 28.9\pm0.6$\,km/s (see Fig.~2a and Methods). Note that $V_{0\alpha}$ is always consistent with $V_{\rm 0B}$ in different schemes, confirming that the $H_\alpha$ emission is indeed associated with the B star-BH binary. The BH mass $M_{\rm BH}$ can then be estimated as $M_{\rm BH}/M_{\rm B} = K_{\rm B}/K_\alpha$, resulting in $M_{\rm BH} = 68^{+11}_{-13}$\,$M_\odot$ (with 90\% errors derived from the measurement uncertainties on $K_B$, $K_\alpha$ and $M_B$). Such a BH mass corresponds to an inclination of $i\approx15^\circ$--$18^\circ$, fully consistent with the wine-bottle shape of the $H_\alpha$ emission line.

The LIGO/Virgo experiments have revealed BHs with masses of several tens of solar masses\cite{Abbott2016,2018arXiv181112907T}, much higher than previously-known Galactic BHs\cite{2006csxs.book..157M, 2014Natur.505..378C}. The discovery of a 70\,M$_\odot$ BH in LB-1 would confirm their existence in our Milky Way. However, while massive stellar BHs are expected to predominantly form in low metallicity (i.e., $<0.2$\,Z$_\odot$) environments\cite{2016Natur.534..512B, 2017NatCo...814906S}, LB-1 has a B-star companion with solar metallicity. This would strongly challenge current stellar evolution models, which only allow for the formation of BHs up to 25\,M$_\odot$ at solar metallicity\cite{2003ApJ...591..288H, Belczynski2010b, 2015MNRAS.451.4086S}. Formation of more massive BHs would require reducing mass loss rates substantially at solar metallicity, and even require overcoming the well-accepted pair-instability pulsations that severely limit BH masses (see Methods).
These strongly-expected limits may suggest that the BH in LB-1 was not formed from the collapse of only one star. One alternative is that LB-1 was initially a triple system, in which the observed B star was the outermost, least massive component, and the present BH was formed by the initial inner binary. Potentially a 70\,M$_\odot$ BH could be formed after a ``normal'' stellar-mass BH merges into the core of a $\gtrsim$60\,M$_\odot$ star during common-envelope evolution, followed by the accretion of the massive star onto its BH core (see Methods). An exciting possibility is that the dark mass still contains two BHs, orbiting each other in an inner binary to which the observed star is a tertiary companion. This requires individual BH masses approaching 35\,M$_\odot$, posing less of a challenge for their formation. In this case, this system would provide a laboratory to test the formation of binary BHs in triple systems.

Our interpretation of an extraordinary 70\,M$_\odot$ dark mass in LB-1 will be undermined if the companion mass is substantially lower than the $8M_\odot$ for the adopted B sub-giant model. 
To accommodate its high luminosity, we need to place the B sub-giant at a distance about twice as large as the 2.14$^{+0.51}_{-0.35}$\,kpc inferred from the Gaia DR2 astrometry\cite{Bailer-Jones2018}. On one hand, this discrepancy could naturally be explained because the binary wobble of the optical component of LB-1 is not accounted for by that Gaia DR2 single-star astrometric solution. In particular, the Gaia DR2 solution shows exceptionally large covariances, suggesting that it is unwise to simply interpret the astrometry as an accurate parallax measurement (see Methods). On the other hand, if LB-1 were indeed at that close distance, with E(B-V) = 0.41\,mag for the appropriate line of sight at that distance\cite{Green2015}, its derived luminosity $L$ would be as low as about 1/6 of the luminosity for the B sub-giant (see Methods). Taking $L \propto M T_{\rm eff}^4/g$, and retaining the same $T_{\rm eff}$ and ${\rm log}g$, this implies a stellar mass $M$ about 1/6 of our adopted value, and consequently a BH mass of about $10M_\odot$. 
No natural stellar models would be consistent with such a companion, but we cannot rule out that the star is in an extreme disequilibrium state (caused for example by a recent outburst or supernova blast from the primary). However, the star should return to equilibrium on the Kelvin-Helmholtz timescale, which for the inferred parameters is about $10^4$ years. Thus, this low mass companion, if true, represents a short-lived disequilibrium phase that would be extremely unlikely to observe.

This wide BH binary shows a surprisingly circular orbit that may shed light on its formation process. Circularization of such a wide binary with tidal torque would take at least a Hubble time, much longer than its age (see Methods). This rules out the possibility that LB-1 was formed by dynamical capture of the B-type star by a BH evolved from a low metallicity star or by a binary BH, as such a capture would result in an eccentric orbit that could not have been circularized by now. In the case of a co-evolving binary, this indicates a very small natal kick along with negligible mass loss when the BH formed. Assuming an initial $e=0$ and a symmetric mass ejection of $\Delta M$ from the BH progenitor, the resultant orbit will have $e = \Delta M / (M_{\rm B} + M_{\rm BH})$. Given that $e = 0.03\pm0.01$, $\Delta M$ must be less than 4\% of the remaining mass, thus helping to form a massive BH. Stellar evolution theories predict fallback supernova and direct BH formation under certain conditions, and some observations might be in favor of their existence, but direct evidence is still lacking despite observational efforts made in the last decade\cite{1999ApJ...522..413F, 2017MNRAS.468.4968A}. LB-1 may be direct evidence for this process.

Unlike every other known stellar BH, LB-1 has not been detected in X-ray observations. We searched for X-ray emission from this system with a 10-kilo-second observation with the Chandra X-ray Observatory, placing an upper limit for the X-ray luminosity of 
$\lesssim 2 \times 10^{31}$\,erg/s (see Methods). This upper limit corresponds to $\sim$10$^{-9}$ of its Eddington luminosity, and suggests a mass accretion rate $\dot{M} \lesssim 10^{-11}$\,M$_\odot$/yr for a conversion efficiency of $\sim$10$^{-4}$ at such low luminosity\cite{1998tbha.conf..148N}. Such low accretion levels could be supplied by the stellar winds of the B sub-giant\cite{deJager1988}. 
Similarly strong $H_\alpha$ emission lines have been observed in some low-mass X-ray binaries in the X-ray quiescent state\cite{1993MNRAS.265..834C, McClintock2003}, where truncated accretion disks do not extend to the innermost BH orbits, thus preventing the emission of measurable X-ray radiation. 
It is long believed that BH binaries in X-ray quiescence can be revealed through radial-velocity monitoring campaigns. The discovery of LB-1, with properties very unlike Galactic BH X-ray binaries, provides such an example. This suggests that future similar campaigns will probe a quiescent BH population different from the X-ray bright one.

\begin{figure*}
\begin{center}
\includegraphics[width=1\textwidth]{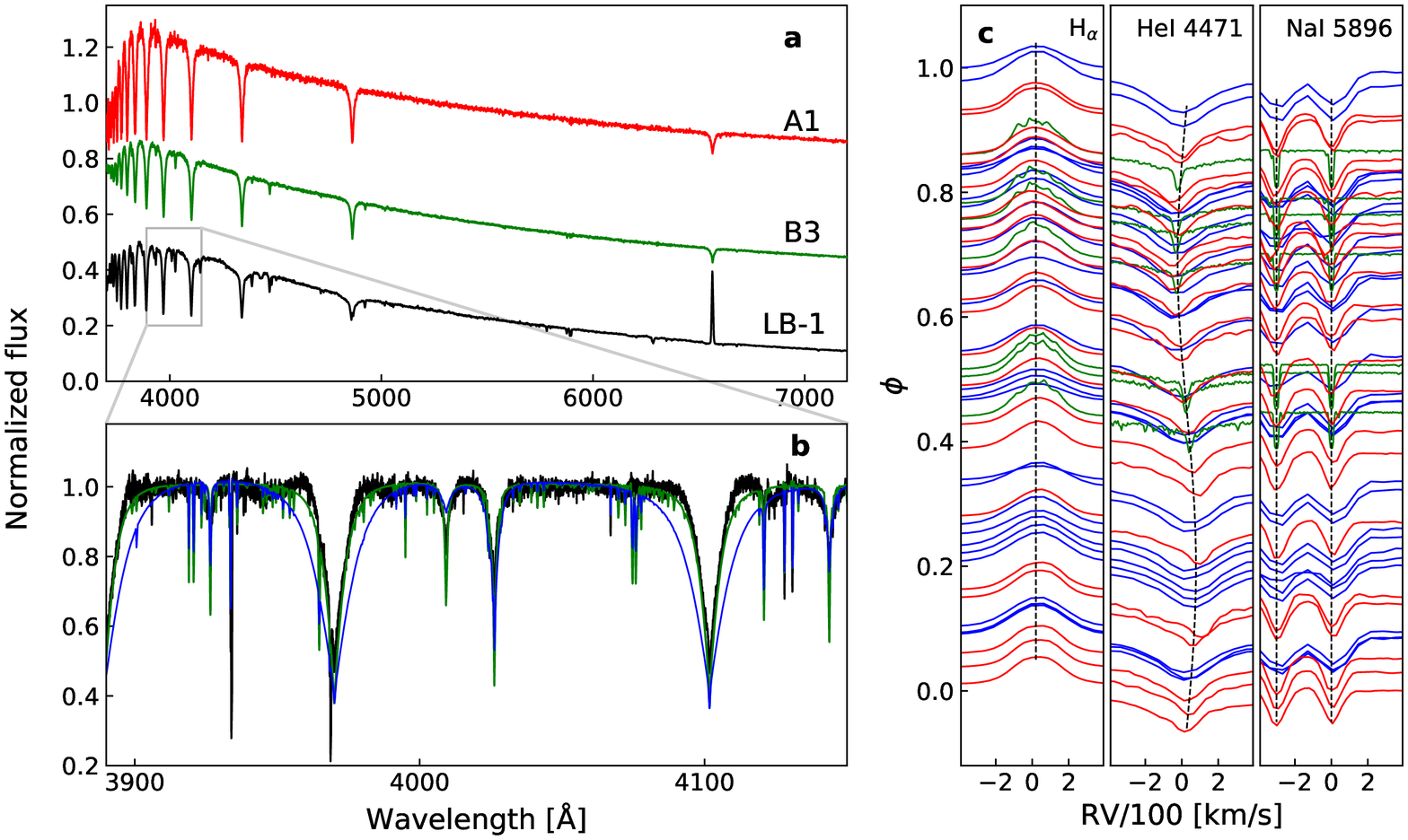}
\caption{{\bf Optical Spectra of LB-1.} {\bf a,} LAMOST spectrum (thin black; $R\approx1,800$) with stellar templates (A1: red; B3: green; offset for clarity) overplotted. {\bf b,} Keck/HIRES spectrum of the wavelength range boxed in {\bf a} (black; $R \approx 60,000$) with the best TLUSTY model (green; $T_{\rm eff} = 18,100$\,K, ${\rm log}g = 3.43$, $Z=$\,Z$_\odot$, $v\sin i = 10$ km/s) overplotted.  The 90\% confidence level (CL) errors for the model are $\Delta T_{\rm eff} = 820$\,K and $\Delta{\rm log}g =0.15$. Also overplotted is a comparison model with ${\rm log}g = 4.75$ (blue), which is the highest ${\rm log}g$ of the model grid but still lower than the typical value (${\rm log}g > 5$) for a B subdwarf. The Balmer absorption lines from this model are much wider than the observed profiles. {\bf c,} Phased line profiles from LAMOST (blue), GTC (red) and Keck (green) observations for $H_\alpha$ emission line, HeI $\lambda4471$ absorption line of the visible star, and interstellar NaI absorption lines. The dashed lines are plotted to guide the eye. The binary phase $\phi$ is for the period of $P=78.9$\,days.}
\end{center}
\end{figure*}

\begin{figure*}
\begin{center}
\includegraphics[width=1\textwidth]{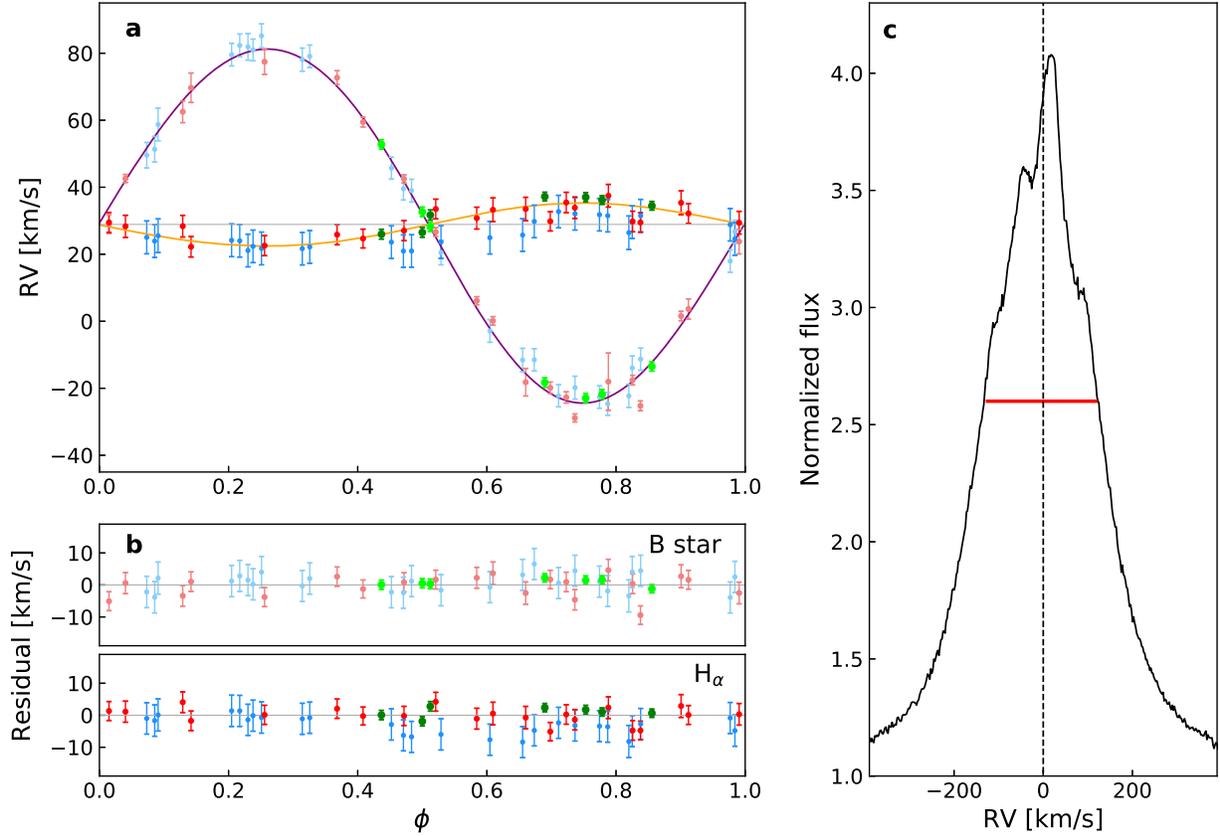}
\caption{{\bf Radial motions of the visible star and the dark primary.} {\bf a,} Folded radial-velocity curves and binary orbital fits for the star and the dark primary as probed by the $H_\alpha$ emission line. The observed data are from LAMOST (blue), GTC (red) and Keck (green). The error bars are the quadratic sum of the wavelength calibration uncertainty and the measurement error. The best-fit binary orbit model for the star (purple) has parameters $K_{\rm B} = 52.8\pm0.7$\,km/s, $e = 0.03\pm0.01$,  and $V_{\rm 0B} = 28.7\pm0.5$\,km/s with a reduced $\chi^2$ of 2.0.  The best-fit model for the $H_\alpha$ emission line (orange) has parameters $K_\alpha = 6.4\pm0.8$\,km/s and $V_{0\alpha} = 28.9\pm0.6$\,km/s with a reduced $\chi^2$ of 0.8. The errors quoted here are for 90\% CL. The gray line with $V_0=28.8$\,km/s is plotted to guide the eye. {\bf b,} Residuals for the binary orbital fits to the star (top) and to the $H_\alpha$ emission line (bottom). The error bars are calculated as above. {\bf c,} Representative $H_\alpha$ emission line profile from one Keck spectrum with high spectral resolution ($R \approx 60,000$). The wine-bottle shape is caused by non-coherent scattering broadening for a disk viewed nearly pole-on. The red line represents a full width at half maximum of about 240\,km/s. }
\end{center}
\end{figure*}

\clearpage

\begin{methods}

\subsection{Discovery and follow up observations of LB-1}
 
LB-1 was among the targets in the LAMOST K2-C0 time domain survey (H.Z. et al., in preparation), which is designed to obtain time domain spectra with LAMOST low resolution spectrograph ($R \approx$ 1,800 over the wavelength range of 3,690--9,100 {\AA}) in a 20 square degree plate chosen from the Kepler K2 Campaign 0. The plate was observed in 26 different nights from 2016 Nov. 7 to 2018 Mar. 23.  The spectra were then reduced with the LAMOST 2D pipeline\cite{Bai2017}. 
 
One aim of the survey is to evaluate the binary mass function $PK^3/2\pi G = {M_{\rm BH}^3 \over (M_{\rm B}+M_{\rm BH})^2} \sin^3i$ once the complete radial-velocity curve can be derived from the time domain spectral data. Here $P$ is orbital period, $K$ is radial-velocity semi-amplitude, and $i$ is viewing angle. Since the mass of the brighter star in the binary can be estimated from its spectrum,  the orbital eccentricity $e$ and radial-velocity semi-amplitude $K$ can  be calculated directly from the radial-velocity curve, then the mass of the dimmer companion can be solved immediately from the mass function given the viewing angle. About two hundreds out of 3,000 target stars turn out to be spectroscopic binaries with periodic radial-velocity variation. Among these, LB-1 exhibits periodic radial-velocity variations with $K$=52.8 km/s, $P$=78.9 days, and $e\simeq$0.

From the relative line strength of HeI $\lambda$4471 vs. MgII $\lambda$4481, we classify the star in LB-1 as a B3V star.
Hot subdwarf B stars (sdBs)  show spectra like B dwarfs but with  much lower mass.
 sdBs have a shorter Balmer series (n$\approx$12), and the HeI $\lambda$4387 is much weaker than HeI $\lambda$4471\cite{SDB}.
In LB-1, the Balmer series extend to more than n$\approx$15, which is at the blue end of LAMOST spectral range, and the HeI $\lambda$4387 is clearly stronger than HeI $\lambda$4471.
In addition, the LAMOST spectra show negligible NII $\lambda$3995 and very weak SiIII $\lambda$4552 lines, which means LB-1 can not be a supergiant (e.g., low mass post-AGB star). 
The information of the LAMOST observations is listed in Extended Data Table 1.

We carried out follow up optical spectroscopic observations of LB-1 with GTC/OSIRIS from 2017 Dec. 2 to 2018 Apr. 26, using the 0$''$.4 slit with three gratings, R2500V, R2500R and R2500I. The spectral coverage for the OSIRIS data is 450--1,000 nm, with a resolution of $\approx$ 3,750.
The spectra were reduced in a standard way with IRAF.
After the bias subtraction and flat correction, the dispersion correction was carried
out based on the line lists given in the manual of OSIRIS
(http://www.gtc.iac.es/instruments/osiris/).
Raw spectra were then extracted with an
aperture size of $\approx$ 6$''$, and a standard star
taken at each night was used to make the flux calibration.
The wavelength calibration uncertainty is about 0.02 \AA~($\approx$ 1.2\,km/s).
The information of the observations is listed in Extended Data Table 1. 

In the period from 2017 Dec. 9 to 2018 Jan. 6, we observed LB-1 on seven individual nights using the Keck I telescope and HIRES spectrometer. Exposure times range from 300 to 600 seconds, and the signal to noise ratio (S/N) per pixel near 550 nm ranges from 80 to 120. Observations were collected using the standard California Planet Search (CPS) setup\cite{Howard2010}, resulting in a spectral resolution of $\approx$ 60,000. The C2 decker (0$''$.87$\times$14$''$.0) was used to allow for removal of night sky line emission features and scattered moonlight. 
We listed the information of the observations in Extended Data Table 1.

\subsection{Stellar Properties from high resolution spectra}

To derive the effective temperature 
($T_{\rm eff}$) and surface gravity (log$g$) of LB-1, we used the spectral libraries BSTAR2006\cite{Lanz2007}, 
which is based on the computer program TLUSTY\cite{Hubeny1995}. The full set of the BSTAR2006 models cover $T_{\rm eff}$ from 15,000 to 30,000\,K with a step of 1,000\,K,
and log$g$ from 1.75 to 4.75 with a step of 0.25\,dex.
These models include six initial metallicities; a micro-turbulence velocity of 2\,km/s is adopted.

The Keck spectra are used to estimate the stellar atmosphere parameters.
Using the lines SiII $\lambda$3856 and SiII $\lambda$5041,
we measured the line width broadened by the stellar rotation.
The program iacob{\_}broad was used in this step, which is available from homepage of the IACOB project (http://research.iac.es/proyecto/iacob/). 
The $v\sin i$ is estimated as $\approx$ 10 km/s, twice that of the spectral resolution (R $\approx$ 60,000) of Keck. 

We performed a rotational and instrumental convolution of the original theoretical
libraries to $v\sin i$ = 10\,km/s and FWHM $=$ 0.1\,{\AA}. Both the theoretical and
observed spectra from the Keck telescope were normalized to a continuum level of unity. We used a Bayesian approach
to estimate the stellar atmosphere parameters, in which each theoretical parameters
is weighed by $e^{-\chi^2/2}$, where $\chi^2$ is the goodness fit of the
model. We obtained a metallicity of 1.18 $\pm$ 0.18\,Z$_{\odot}$ with SiII $\lambda$4131 and 
1.17$\pm$0.11\,Z$_{\odot}$ with MgII $\lambda$4481. 
Finally, using the theoretical grids with solar abundances and the observed hydrogen lines in range of 3750--4150\,{\AA},
we obtained $T_{\rm eff}$ as 18,104 $\pm$ 825\,K and log$g$ as 3.43 $\pm$ 0.15\,dex. 
The errors were estimated using the standard deviations of the fitting results from the seven Keck spectra.
These parameters prove that the optical counterpart is a B star.

Assuming the solar metallicity ($Z = 0.017$),
we determined the mass, radius, and age of the B star.
The evolutionary grid of $T_{\rm eff}$ and log$g$ for stars with different initial masses 
were constructed based on the PARSEC isochrones\cite{Bressan2012,Marigo2017} (Downloaded from http://stev.oapd.inaf.it/cgi-bin/cmd\_3.1.). 
In the Extended Data Figure 1, stars located in the ellipse are considered as acceptable for the B star.
We downloaded the sequences of isochrones at small steps of $\Delta(\log~t) = 0.0025$, and 
collected the points inside the ellipse as acceptable models.
Finally, we find that at $Z = 0.017$, the physical solutions (with 90\% uncertainty) consistent with our constraints are:
$M_{\rm B} = 8.2^{+0.9}_{-1.2}$\,M$_{\odot}$; 
$R_{\rm B} = 9\pm2$\,R$_{\odot}$;
$t_{\rm age} = 35^{+13}_{-7}$\,Myr.

\subsection{Distance and interstellar extinction}

The spectral energy distribution (SED) of LB-1 
was extracted from the UCAC4 catalog, 2MASS and AllWISE data release. We used the acceptable PARSEC models to construct a grid of SEDs. By comparing the observed SEDs with the PARSEC ones, 
we fitted the distance and $E(B-V)$ simultaneously.
Considering that the accretion disk and circumbinary materials can result in radiation in the near- and mid-infrared bands\cite{Bonatto06},
only the $U$, $B$, $V$ magnitudes were used in the fitting.
We presented the fitting results in Extended Data Figure 2. The excesses can be found from $K_{\rm S}$ to W4 bands. 
The best fit yields the reddening value as $E$($B-V$) $=$ 0.55 $\pm$ 0.03 mag and the distance as 4.23 $\pm$ 0.24 kpc (with 90\% uncertainty). 
For such a distance, the Pan-STARRS 3D dust map returns an extinction of $E$($B-V$) $\approx$ 0.6, consistent with our fitting result (Extended Data Figure 3).

The distance derived above is larger than the  $2.14^{+0.51}_{-0.35}$ kpc value from the Gaia data release 2 catalog\cite{Bailer-Jones2018}. This is possibly because the $Gaia$ DR2 solution has assumed a single star for LB-1, and has mistaken the binary motion itself as part of the parallax,
making the parallax and distance unreliable.
In Gaia DR2, the covariances between position parameters (ra, dec) and parallax of LB-1 are much higher than other sources. The covariance dec\_parallax\_corr from the Gaia DR2 is $-$0.62, higher than the absolute value of dec\_parallax\_corr of 98\% sources between 10 and 13 mag ($G$ band). The covariance of ra\_parallax\_corr is 0.54 which is also higher than that of 96\% sources in the magnitude range.

Given the mode of operation of the Gaia astrometric instrument and the actual along scan single-CCD measurement error of 0.3 mas\cite{Lindegren2018}, an astrometric error as small as 0.1 mas per visit can be anticipated for the 11.5 magnitude of LB-1 (in each visit a star is measured on up to 9 different astrometric CCDs). Given its location, a total of about 80 of such visits are predicted throughout the Gaia 5-yr nominal mission lifetime (ending in September 2019). With these numbers in mind, there is actual hope that the 0.4 mas astrometric orbital motion can be uncovered once the single vist data are properly reduced and/or made available in the future.

The hot sdB scenario can also be rejected from the distance. The Gaia DR2 catalogue shows hot subluminous stars have an absolute magnitude around 5 mag\cite{Geier2019}. With the $G$-band magnitude of 11.918 mag for LB-1, the distance estimation for a sdB would be less than 240 pc. This is seriously inconsistent with the Gaia DR2 distance and our fitting result, also inconsistent with the clear diffuse interstellar bands (DiBs)\cite{Friedman2011} in the spectra which should be much shallower for a sdB at this distance.

As a test, we calculated the radius and mass of the B star using the observational parameters, including the $V$-band magnitude ($\approx$ 11.51\,mag), the reddening value, the distance, the effective temperature, and the surface gravity.
With the bolometric correction\cite{Girardi2004} being $\approx -$1.6, the bolometric magnitude of the B star is $M_{\rm B, bol} = -4.93$ mag, and the bolometric luminosity is calculated as 
$L_{\rm B,bol} = L_{\rm \odot,bol} \times 10^{0.4(M_{\rm \odot, bol} - M_{\rm B, bol})} \approx 7,000$\, L$_{\rm \odot, bol}$.
The solar bolometric magnitude and luminosity are 4.74 mag and $3.828\times10^{33}$ erg/s, respectively.
The radius is calculated as 
$R_{\rm B} = \sqrt[]{\frac{L_{\rm B, bol}}{4 \pi \sigma T^4}} \approx$ 8.7\,R$_{\odot}$,
and the mass is calculated as $M_{\rm B} = \frac{gR_{\rm B}^2}{G}  \approx 7.5$\,M$_{\odot}$.
Both of them are consistent with the PARSEC model fitting results.
The Kelvin-Helmholtz Timescale $t_{\rm KH} = \frac {GM^2}{RL}$ is defined as the time required to radiate current gravitational binding energy at its current luminosity, and represents the timescale for a star in disequilibrium to adjust back to equilibrium. In our case, $t_{\rm KH}$ is around $2.7\times10^4$ yr.

However, if we use the Gaia distance ($\approx$ 2.14 kpc), which corresponds to an extinction of $E(B-V) = 0.41$ from Pan-STARRS 3D dust map, the bolometric luminosity can be estimated as $L_{\rm B,bol} \approx$ 1,300\,L$_{\rm \odot, bol}$. The radius and the mass would be $R_{\rm B} \approx$ 3.6\,R$_{\odot}$ and $M_{\rm B} \approx 1.3$\,M$_{\odot}$, respectively. Using these parameters, the Kelvin-Helmholtz Timescale would be $t_{\rm KH} \approx 1.1\times10^4$ yr.

Furthermore, if we use the Gaia distance ($\approx$ 2.14 kpc)
and assume an extinction of $E(B-V) = 0.55$ as derived from fitting the B subgiant model to the spectral energy distribution, the bolometric luminosity can be estimated as $L_{\rm B,bol} \approx$ 1,900\,L$_{\rm \odot, bol}$. The radius and the mass would be $R_{\rm B} \approx$ 4.4\,R$_{\odot}$ and $M_{\rm B} \approx 1.9$\,M$_{\odot}$, respectively. The Kelvin-Helmholtz Timescale is then estimated as $t_{\rm KH} \approx 1.4\times10^4$ yr.

We conclude that if we place the companion at the Gaia DR2 distance, with $E(B-V)$ ranging from 0.41\,mag to 0.55\,mag, we will get a star in disequilibrium, with the Kelvin-Helmholtz timescale of 11,000--14,000 years.

\subsection{Radial velocity measurements}
For the B star, we measured the radial velocity by matching the model templates using the cross correlation method. 
For LAMOST data, we removed the Balmer lines and DiBs, and fitted the  spectrum ranging from 4,000 to 5,200 \AA.
For GTC and Keck data, we removed DiB and used the spectrum ranging from 4,000 to 6,000 \AA.

Firstly, we Doppler shifted the best theoretical spectra to a set of radial velocities. 
Secondly, we calculated the $\chi^2$ by comparing these model spectra with those observed ones, and used the radial velocity with minimum $\chi^2$ as the best estimation.
Also, we calculated the systematic shifts between these exposures by comparing the absorption band of water vapor in range of 6,850--6,940\,{\AA}. We first used the first exposure as the reference spectra, and calculated the radial-velocity shift by cross correlating it with the other exposures. Then we used the exposure with the median value of shift as new reference spectra, performed the calculation again, and obtained the final systematic shifts.

One key step before radial velocity measurements is to justify whether the $H_\alpha$ emission is around the BH or from a circumbinary disk.
For a circumbinary disk, the Keplerian velocity is $\sqrt[]{G(M_{\rm BH}+M_{\rm B})/1.7a}$, where $a$ is the binary separation and 1.7$a$ is the typical inner radius\cite{Artymowicz1994}. 
The velocity of the visible star is $\sqrt[]{G(M_{\rm BH}+M_{\rm B})/a_{\rm B}}$, where $a_{\rm B}$ is the distance from the visible star to the barycenter ($a_{\rm B}$ $=\frac{M_{\rm BH}}{M_{\rm B}+M_{\rm BH}} a$). 
The projected velocity at the inner radius of the circumbinary disk would be $\approx {1 \over \sqrt{1.7}} \approx 0.75$ times that of the visible star (52.8 km/s), i.e., about 40 km/s.
However, the observed line is three times wider with an FWHM of 240 km/s.
This means that the $H_\alpha$ emission line comes from a disk around the BH rather than a circumbinary disk.

While it is clear that the H$\alpha$ emission line is associated with the BH, it is tricky to track the BH motion through the H$\alpha$ line, because the complex structures in the line center may be contaminated by components such as circumbinary materials, gravitationally focused accretion streams, and hot spots in the accretion disk. These components are not symmetrically centered on the BH, hence their motion will not be in exact phase with that of the BH disk,and will act to decrease the BH motion if we include them in the calculation. Note that the line profiles can not be fitted with simple analytic forms such as Gaussian or Lorentzian profiles, so we decided instead to infer radial velocities using the barycenter of the line. 

First we calculate the barycenter of the whole H$\alpha$ profile. The derived radial velocities over two years can be folded with the orbital period, resulting in a sinusoid with an amplitude of $1.7\pm0.9$ km/s in anti-phase with the B star velocity. Such a line velocity, if it should represent the BH motion, would suggest a mass ratio of 20--67 given $M_{\rm BH}/M_{\rm B} = K_{\rm B}/K_\alpha$, hence a BH mass of 140--600\,$M_\odot$. Second we mask out the core of the line profile and calculate the barycenter from the unmasked line wings. We start by measuring the barycenter from velocity bands constrained between 1/2 FWHM up to 500 km/s, on each side of the line profile. This results in $K_\alpha = 4.4 \pm 0.7$ km/s, as shown in Extended Data Table 2. This demonstrates that the line center is indeed contaminated by components not centered on BH that will act to decrease the measured BH motion.

To explore the systematics of the derived BH motion, we experiment with different mask limits with inner edges in the range corresponding to from 2/3 to 1/5 heights of the H$\alpha$ emission line, and inner edges at 120/140/170/200 km/s from the barycenter. The resulting amplitudes vary between $3.9\pm0.8$ km/s and $6.7\pm1.0$ km/s as summarized in Extended Data Table 2. It is clear from the table that the anti-phased H$\alpha$ motion has larger amplitudes as we move away from the central part of the line, but begin to saturate after 1/3 height. This suggests that the unmasked line wings outside 1/3 height can largely avoid contamination from the line center, and we decide to use the 1/3 height masking scheme to represent the BH motion, i.e.,  $K_\alpha = 6.4 \pm 0.8$ km/s as shown in Table 2 and Table 3 again. 

If we had many more high resolution Keck/HIRES spectra covering all binary phases, we would be able to reconstruct the morphology of the accretion disk and other components, giving a detailed description of its asymmetric shape and sizes. This of course will give us a more accurate determination of the BH mass, and we will pursue such a (costly) follow-up campaign in the coming years. Our current LAMOST/GTC/Keck observations, however, are enough for a rough estimate of the BH mass already.

For LAMOST and GTC observations, there are multiple exposures during a single night. 
We used the averaged value as the radial velocity at that day.
The measurement error was estimated using the standard deviation of multi-exposures during one night.
For Keck observations, we use the measurement error, which is about 1\,km/s.
The system difference between days are calibrated by both the telluric emission (for LAMOST) or absorption(for GTC and Keck) lines and the diffuse interstellar absorption lines/bands  at NaID lines, 5782\AA\  and 6284\AA.
The uncertainty for the radial velocity is the quadratic sum of the wavelength calibration uncertainty and the measurement error.

\subsection{Period and orbital parameters}

Using the Lomb-Scargle\cite{Lomb1976, Scargle1982} method, we measured the period of LB-1 with the radial-velocity curve from LAMOST, GTC, and Keck observations. The period is 78.9$\pm$0.3\,day (Extended Data Figure 4).
We fitted the radial velocity data of the B star (54 points) and H$\alpha$ line wing (54 points) simultaneously,
using the equation 
\begin{equation}
V=K[\cos (\theta+\omega)+e\cos(\omega)]+V_0,
\end{equation}
where $K$ is the simi-amplitude of the radial velocity curve, $\theta$ is the phase angle, $\omega$ is the longitude of periastron, and $V_0$ is the system velocity.
The best fit parameters (i.e., eccentricity $e$, semi-amplitude $K_{\rm B}$ and $K_{\rm \alpha}$, velocity $V_{\rm 0B}$ and $V_{\rm 0\alpha}$) are listed in Extended Data Table 3.
The best-fit for the B star motion has a reduced $\chi^2$ of 2.0, while the best-fit for the $H_\alpha$ motion (for the 1/3 height scheme) has a reduced $\chi^2$ of 0.8.
To obtain the uncertainty of one parameter, we fixed other parameters at the best fit values and re-did the fitting. Then, the uncertainty of that parameter was estimated with $\Delta~\chi^2 = $ 2.706 (in 90\% confidence) and $\Delta~\chi^2 = $ 6.635 (in 99\% confidence), respectively.

We compared the fittings of the H$\alpha$ velocity using one sinusoid and one horizontal line. For the sinusoid fitting, there are two free parameters (i.e., $K_{\rm \alpha}$ and $V_{\rm 0\alpha}$); for the line fitting, there is one free parameter (i.e., $V_{\rm 0\alpha}$). Therefore, the degree of freedom for the two fittings are 52 and 53, respectively. The $\chi^2$ for the two fittings are 109.49 and 219.24, respectively. Using the F-test, we find the sinusoid fitting is statistically significantly better than the line fitting (P $<$ 0.01\%).

The separation $a$ can be calculated from the Kepler's Third Law $a=[\frac{G(M_{\rm B}+M_{\rm BH})P^2}{4\pi^2}]^{\frac{1}{3}}$ for each pair of $M_{\rm BH}$ and $M_{\rm B}$. 
The ranges of the separation $a$ and the semi-major axis $a_{\rm B}$ are shown in Extended Data Figures 5 and 6 respectively under the limitations of $M_{\rm B}$ and $M_{\rm BH}$,
which clearly show LB-1 is a wide binary.

\subsection{BH formation}
\subsection{Individual stellar progenitor scenario} 
First, let us assume that the dark object in LB-1 is a single BH formed from an individual star. Its mass depends on three major factors: (i) initial stellar mass; (ii) wind mass loss during the star's life; (iii) BH formation process during the final core-collapse/supernova. The initial stellar mass sets an upper limit to the BH mass, while winds and collapse/explosion processes are responsible for removing stellar mass and reducing the BH mass. All these aspects of stellar evolution are highly uncertain, which allows for a wide range of possibilities when it comes to BH mass calculations.

Guided by observations (or the lack thereof), we constructed a set of models based on stellar evolution calculations\cite{Hurley2000} to estimate the maximum BH mass at solar metallicity ($Z=0.017$). We allow stars to form with initial masses as high as $200$ M$_\odot$; at least one such star has already been discovered\cite{Crowther2010}. 
Recent observations indicate that stellar winds may be overestimated by as much as a factor of $10$ for some massive stars\cite{Rama2019}, compared with standard values\cite{Vink2001}; hence, in our calculation we reduce the theoretically predicted wind mass-loss rates by a factor of 2 to 3. 
At the end of a massive star's life, we allow for direct BH formation with no supernova explosion or associated mass loss\cite{1999ApJ...522..413F}. Such mode of BH formation is supported by the low peculiar velocities observed in the most massive Galactic BHs known to-date\cite{Mirabel2003}, and by the claimed observation of a luminous star disappearing without a supernova\cite{2017MNRAS.468.4968A}. Finally, we also eliminate mass loss from pair-instability pulsations during the supernova explosion; we note that, despite the large amount of theoretical work on pair-instability mass loss and supernovae, so far there is no observational evidence to support this mechanism. 

We have incorporated all these options into the population synthesis code {\tt StarTrack}\cite{Belczynski2008,Belczynski2017} to estimate the maximum BH mass in the Galaxy.  
In Extended Data Figure 7, we present our models that challenge the currently accepted paradigms, but that can possibly explain the BH mass in LB-1.  
Our first model (magenta line) shows the standard prediction of BH masses as a function of initial stellar masses. BHs form only with relatively low masses of $\approx$ 5--15 M$_\odot$, as a result of strong stellar winds that remove most of the stellar mass before core-collapse. 
In our second model (blue line), we reduce stellar winds by a factor of $2$. This reduction factor is applied to all types of winds: from O stars, B supergiants, luminous blue variables, and Wolf-Rayet stars\cite{Belczynski2010b}. As a result, the BH mass can reach $\approx$30 M$_\odot$. 
In our third model (red line), we reduce winds by a factor of $3$. 
The maximum BH mass is now $\approx$60 M$_\odot$, from a star with an initial mass of 120 M$_\odot$, which loses half of its mass in stellar winds before direct collapse. Stars more massive than $120$ M$_\odot$ grow massive Helium cores ($M_{\rm He}>45$ M$_\odot$) and are thus subject to pair-instability pulsation supernova mass losses. Precise estimates of this type of mass loss are model-dependent, but most models agree that the BH remnants are less massive than $\approx$50 M$_\odot$\cite{Woosley2017,Leung2019}. In our model, BHs formed after pair-instability pulsations are assumed to be always less massive than $\approx$40 M$_\odot$\cite{Belczynski2016a}.  
In our fourth model (black line), we not only reduce stellar winds by a factor of $3$, but also turn-off pair-instability pulsation supernova mass losses.
The maximum BH mass reaches $\approx$80 M$_\odot$, for a maximum initial stellar mass of 200 M$_\odot$: enough to explain the dark mass in LB-1 as a single stellar BH.

\subsection{Binary progenitor scenario} 
Let us suppose now that the progenitor of LB-1 consisted of a massive (but not extraordinary) binary, with two stars of initial mass $\gtrsim$60 M$_\odot$ each, and a much less massive third star  (the B3 star we see today) orbiting around the O-star pair. The more massive of the two O stars evolves first, forming a BH with a mass $\approx$10--20 M$_\odot$ at solar metallicity. If the other O star has a mass $\gtrsim$3.5 times the BH mass, the system is thought to evolve through a common envelope phase\cite{2016Natur.534..512B,Dominik2012,vandenHeuvel2017}. Let us then assume that the BH sinks towards the core of the O star before the common envelope is ejected. What happens at this stage is an open question; one scenario is that the core is tidally disrupted and accreted by the inspiralling BH, in a regime of radiatively inefficient (advective), hyper-critical accretion\cite{Jiang2014,Sadowski2014,Abramowicz2013}. If the radiative and mechanical feedback from the accreting BH is not sufficient to destroy the star, or is collimated along the polar direction, most of the O-star envelope may end up also being accreted into the BH core\cite{Abubekerov2009}, in a kind of triggered direct collapse. The final result may be a single BH with a mass $>$60 M$_{\odot}$ with the B3 star still orbiting around it.

An alternative scenario is that the O-star plus BH binary system is too wide to undergo a common envelope phase, or the mass ratio is not high enough, and the system evolves instead in a slower spiral-in\cite{vandenHeuvel2017}. At the end of this phase, the O star will also collapse into a BH, without a merger. The dark mass measured in LB-1 may be a binary BH, with masses of $\approx$35 M$_\odot$ for each component. The advantage of this scenario is that the formation of two 35\,M$_\odot$ BHs from two massive stars is less problematic than the formation of a 70\,M$_\odot$ BH from a single star. In this scenario, too, the B3 optical counterpart is the small third component of the triple stellar system.

\subsection{Circularization timescale}

Tidal interactions in a binary tend to circularize its initially eccentric orbit.
To estimate the circularization timescale of a B-BH binary, we use the MESA code\cite{Paxton2011} to simulate the orbital evolution. The B star mass and the orbital period are initially set to be 8 M$_{\odot}$ and 79 days, respectively. During the evolution, we follow the evolution of the B star from the zero-age main-sequence phase to the age of 50\,Myr when the star slightly evolves off the main-sequence stage. Since such a B star has a radiative envelope, we adopt the mechanism involving dynamical tides with radiative damping\cite{h02} to deal with the binary orbital evolution.

We vary the initial orbital eccentricity in the range of 0.1--0.5 and the initial BH mass in the range of 40--1,000\,M$_{\odot}$ to test their influence on orbital circularization. Our calculations show that the orbital eccentricity of the binary is nearly unchanged and the corresponding circularization timescale is always larger than 10$^{14}$\,years over the whole 50\,Myr. It is argued that a significant enhancement of radiative damping is required to match the observed eccentricity-period distribution in late-type binaries\cite{Claret2007}, so our calculated circularization timescale may be overestimated in some extent. Since the B star (with a radius less than its current value of $9\pm2$\,R$_\odot$) is well within its Roche lobe (with a size of 73--71\,R$_\odot$ corresponding to the BH mass of 10--100\,M$_{\odot}$), tides are not expected to be important independent of the mechanism behind tidal damping.

If decreasing the masses of both components of the binary system by a factor of 6, the BH's companion is now a low-mass ($\approx$1.3 M$_\odot$) star with a convective envelope. We then apply the mechanism involving equilibrium tide with convective damping\cite{h02} to simulate the binary orbital evolution. We find that the circularization timescale is
still larger than 10$^{12}$\,years before the low-mass star climbs to the red giant branch (corresponding to the radius of $\approx$3\,R$_\odot$).

\subsection{X-ray luminosity and Eddington ratio.}

We obtained a 10-ks DDT observation with {\it Chandra} ACIS-S3 on 2018 Jan 13. We reprocessed the data with {\sc ciao} version 4.10; we used the {\sc ciao} task {\it {scrflux}} for flux measurements. We do not detect the source, which places a 90\% upper limit to the 0.5--7 keV net count rate of $\approx$3.8 $\times 10^{-4}$ ct s$^{-1}$.

In order to convert this limit to an unabsorbed flux limit, we used a grid of plausible values of photon index and column density. The typical power-law photon index of black hole binaries in the quiescent state is $\Gamma \sim 1.5$--2.1\cite{Plotkin2013,RM2006}. To constrain the column density, we used the best-fitting value of the optical reddening E(B-V) = 0.55 mag. Applying the standard linear relation between the hydrogen column density $N_{\rm H}$ and the reddening\cite{Schlegel1998} $N_{\rm H} = 5.8 \times 10^{21}/E(B-V)$, we obtain $N_{\rm H} \approx 3.2 \times 10^{21}$ cm$^{-2}$. A similar result ($N_{\rm H} \approx (3.1$--$3.8) \times 10^{21}$ cm$^{-2}$) is obtained from the best-fitting relation between $A_V \equiv 3.1 E(B-V)$ and hydrogen column density\cite{Predehl1995,Guver2009}. The line-of-sight Galactic column density in the direction of LB-1 provides a plausible upper limit\cite{Kalberla2005} $N_{\rm H} \approx 4.7 \times 10^{21}$ cm$^{-2}$. The saturated relation\cite{Liszt2014} provides a lower limit $N_{\rm H} \approx 1.0 \times 10^{21}$ cm$^{-2}$ for $E(B-V) = 0.55$ mag.  The result of our analysis over this range of photon indices and column densities is that LB-1 is not detected down to a 90\% upper limit of $f_{0.3-8} < 3.9 \times 10^{-15}$ erg cm$^{-2}$ s$^{-1}$ for the absorbed flux in the 0.3--8 keV band (assuming the softest slope), or $f_{0.3-8} <  4.8 \times 10^{-15}$ erg cm$^{-2}$ s$^{-1}$ (assuming the hardest slope). At the adopted distance of 4.23 kpc, the 90\% upper limits for the emitted luminosity are $L_{0.3-8} < 1.2 \times 10^{31}$ erg s$^{-1}$ (assuming the lowest limit of $N_{\rm H}$), or $L_{0.3-8} < 1.8 \times 10^{31}$ erg s$^{-1}$ (assuming the highest value of $N_{\rm H}$). Finally, for our inferred BH mass of $\approx$70 $M_{\odot}$, this corresponds to an Eddington ratio $L_{\rm X}/L_{\rm Edd} \lesssim 2 \times 10^{-9}$. This is the lowest value recorded for a quiescent Galactic BH binary\cite{Garcia2001,McClintock2004,Plotkin2013}, and similar or lower than in any quiescent nuclear BH in nearby galaxies\cite{Yuan2009,Ho2009,Gallo2010}.

At very low accretion rates, the radiative efficiency $\eta$ is reduced: a standard scaling for the ADAF model\cite{Narayan1998,Yuan2014} is $\eta \sim 10 \dot{m}$ where $\dot{m} \equiv \dot{M}/\dot{M}_{\rm {Edd}}$ and $\dot{M}_{\rm {Edd}} \equiv L_{\rm{Edd}}/(0.1 c^2)$. 
Here $\dot{M}$ is the accretion rate, $\dot{M}_{\rm {Edd}}$ is the Eddington accretion rate, and $\dot{m}$ is the Eddington ratio.
A similar scaling of $\eta \approx 0.7 \, (\alpha/0.3) \, (L/L_{\rm{Edd}})^{1/2}$ was derived\cite{Mahadevan1997}. 
An even steeper dependence of $\eta$ with accretion rate ($\eta \propto \dot{m}^{1.3}$, $L \propto \dot{m}^{2.3}$) was proposed\cite{Merloni2003,Russell2013}. Thus, our observed Eddington ratio $L_{\rm X}/L_{\rm Edd} \lesssim 2 \times 10^{-9}$ suggests $\dot{M} \lesssim 10^{-5} \dot{M}_{\rm {Edd}} \approx 10^{-11}  M_{\odot}$ yr$^{-1}$ (with an uncertainty of a factor of 2, between alternative scaling approximations of the radiative efficiency at low accretion rates).

\end{methods}

\begin{addendum}

\item[Acknowledgements]

We thank Drs. Daniel Wang, Jon Miller, Ed Cackett, Ramesh Narayan, Hailiang Chen, Bing Zhang, Christian Motch, Mike Bessel, Gary Da Costa, Alexey Bogomazov, Songhu Wang, and many others for helpful discussions. 
This work is supported by the National Science Foundation of China (NSFC) under grant numbers 11988101/11425313 (J.L.), 11773015/11333004/U1838201 (X.L.), 11603010 (Y.S.), 11690024 (Y.Lu), U1531118 (W.Z.), 11603035 (S.W.), 11733009 (Q.L.), and 11325313/11633002 (X.W.). It is also supported by the National Key Research and Development
Program of China (NKRDPC) under grant numbers 2016YFA0400804 (J.L.), 2016YFA0400803 (X.L.), and 2016YFA0400704 (Y.Lu).
JC acknowledges support by the Spanish Ministry of
Economy, Industry and Competitiveness (MINECO) under grant AYA2017-83216-P.
KB acknowledges support from the Polish National Science Center (NCN) grants OPUS (2015/19/B/ST9/01099) and Maestro (2018/30/A/ST9/00050).
This work is only made possible with LAMOST (Large Sky Area Multi-Object Fiber Spectroscopic Telescope), a National Major Scientific Project built by the Chinese Academy of Sciences. Funding for the project has been provided by the National Development and Reform Commission. LAMOST is operated and managed by the National Astronomical Observatories, Chinese Academy of Sciences. 
This work is partly based on observations made with the Gran Telescopio Canarias (GTC), installed in the Spanish Observatorio del Roque de los Muchachos of the Instituto de Astrofísica de Canarias, in the island of La Palma.	
Part of the data was obtained at the W.M. Keck Observatory, which is operated as a scientific partnership among the California Institute of Technology, the University of California and the National Aeronautics and Space Administration. The Observatory was made possible by the generous financial support of the W.M. Keck Foundation. 
The scientific results reported in this article are based in part on observations made by the Chandra X-ray Observatory (ObsID 20928). This research has made use of software provided by the Chandra X-ray Center (CXC) in the application packages CIAO.

\item[Author contributions]

J.L. and H.Z. are equally responsible for supervising the discovery and follow-up observations. H.Z. and Z.H. proposed the LAMOST monitoring campaign, and H.Z.'s group reduced the LAMOST data with meticulous efforts. J.L. proposed the GTC/Keck/Chandra observations, and his and H.Z.'s groups carried out subsequent data reduction and analysis. J.L. wrote the manuscript with help mainly from H.Z., Y.Lu, R.S., S.W., X.L., Y.S., T.W., Y.B., Z.B., W.Z., Q.G., Y.W., Z.Z., K.B. and J.C. 
W.W., A.H., W.M.G., J.Wang, J.Wu, L.S., R.S., X.W., J.B., R.D.S. and Q.L. also contributed to the physical interpretation and discussion. 
H.Y., Y.D., Y.Lei, Z.N., K.C., C.Z., X.M., L.Z., T.Z., H.W., J.R., Junbo Zhang, Jujia Zhang and X.W. also contributed to data collection and reduction.
A.M.H and H.I. contributed to collecting and reducing Keck data.
A.C.L., R.C. and R.R. contributed to collecting and reducing GTC data.
Z.Q., S.L. and M.L. contributed to utilization of Gaia data.
Y.Z., G.Z., Y.C. and X.C. contributed to the implementation of LAMOST.
All contributed to the paper in various forms.

\item[Author information] 

Reprints and permissions information is available at npg.nature.com/reprints.  The authors declare that they have no competing financial interests.  Readers are welcome to comment on the online version of the paper.  Correspondence and requests for materials should be addressed to J.L. (email: jfliu@nao.cas.cn) and/or H.Z. (email: htzhang@bao.ac.cn).

\item[Data availability]

The data that support the plots within this paper and other findings of this study are available from the corresponding authors upon reasonable request.

\end{addendum}

\clearpage
\newpage

\begin{table}
\vspace{-2cm}
\label{lamost.tab}
\caption*{\bf Extended Data Table 1. Spectral Observations of LB-1.}
\scriptsize
\begin{center}
\renewcommand\arraystretch{0.8}
\begin{tabular}{lccccc}
\hline
Instrument & Date        &Exposure Time & Phase  &  $RV_{\rm B}$ & $RV_{\rm \alpha}^a$\\
   &         &        (second)  &   &   (km/s)   & (km/s)\\
(1)         & (2)  &  (3)  & (4)   &  (5) & (6)\\
\hline
\multirow{26}{0.5in}{LAMOST} & 2016.11.07    & 600$\times$15&0.47&39.5$\pm$3.4&20.9$\pm$4.9\\
& 2016.11.08    & 600$\times$11&0.48&39.0$\pm$3.4&21.0$\pm$4.9\\
& 2016.11.23  &600$\times$12&0.67&$-$11.5$\pm$3.3&29.8$\pm$4.9\\
& 2016.11.26  &600$\times$13&0.71&$-$22.2$\pm$3.3&32.7$\pm$4.9\\
& 2016.11.28  &600$\times$13&0.74&$-$19.8$\pm$3.4&32.1$\pm$4.8\\
& 2016.12.01  &600$\times$14&0.77&$-$22.6$\pm$3.3&31.8$\pm$4.8\\
& 2016.12.02  &600$\times$14&0.79&$-$24.6$\pm$3.4&31.5$\pm$4.8\\
& 2016.12.05  &600$\times$12&0.83&$-$14.0$\pm$3.6&29.6$\pm$4.8\\
& 2016.12.06   &600$\times$7&0.84&$-$11.3$\pm$3.3&31.5$\pm$4.9\\
& 2016.12.17    & 600$\times$13&0.98&18.0$\pm$3.4&28.8$\pm$4.9\\
& 2016.12.26    & 600$\times$9&0.09&58.7$\pm$4.9&25.5$\pm$5.0\\
& 2017.01.04    & 600$\times$8&0.20&79.6$\pm$3.4&24.2$\pm$4.9\\
& 2017.01.05    & 600$\times$8&0.22&82.3$\pm$3.4&24.0$\pm$4.9\\
& 2017.01.06    & 600$\times$7&0.23&81.9$\pm$4.0&21.1$\pm$4.9\\
& 2017.11.18    & 600$\times$8&0.24&81.0$\pm$3.2&22.4$\pm$4.9\\
& 2017.11.19    & 600$\times$8&0.25&85.2$\pm$3.6&21.7$\pm$4.9\\
& 2017.11.24    & 600$\times$10&0.31&78.1$\pm$3.4&21.7$\pm$4.9\\
& 2017.12.11    & 600$\times$11&0.53&20.5$\pm$3.7&23.7$\pm$4.9\\
& 2017.12.17    &600$\times$8&0.60&$-$2.9$\pm$3.4&24.9$\pm$4.9\\
& 2017.12.21   &600$\times$8&0.66&$-$11.6$\pm$3.5&25.7$\pm$4.9\\
& 2018.01.16    & 600$\times$8&0.98&26.7$\pm$3.4&24.6$\pm$4.9\\
& 2018.01.23    & 600$\times$9&0.07&49.6$\pm$3.8&25.0$\pm$4.9\\
& 2018.01.24    & 600$\times$8&0.09&51.3$\pm$3.8&23.9$\pm$4.9\\
& 2018.02.12    & 600$\times$8&0.33&79.1$\pm$3.4&22.2$\pm$4.9\\
& 2018.02.22    & 600$\times$7&0.45&45.7$\pm$3.3&23.6$\pm$4.9\\
& 2018.03.23   &600$\times$3&0.82&$-$22.3$\pm$3.4&26.4$\pm$5.0\\
\hline 
\multirow{21}{0.5in}{GTC} & 2017.12.02   & V 30$\times$3, R 30$\times$3, I 30$\times$3 &0.41&59.5$\pm$1.5&24.7$\pm$2.8\\
&2017.12.07   & V 30$\times$3, R 30$\times$3, I 30$\times$3 &0.47&42.5$\pm$1.2&27.1$\pm$3.0\\
&2017.12.10   & V 30$\times$3, R 30$\times$3, I 30$\times$3 &0.52&26.6$\pm$1.3&33.5$\pm$2.9\\
&2017.12.17   & V 30$\times$3, R 30$\times$3, I 30$\times$3 &0.61&0.1$\pm$1.3&33.3$\pm$3.5\\
&2017.12.21   & V 30$\times$3, R 30$\times$3, I 30$\times$3 &0.66&$-$18.2$\pm$4.1&33.5$\pm$3.5\\
&2017.12.26   & V 30$\times$3, R 30$\times$3, I 30$\times$3 &0.72&$-$22.7$\pm$1.7&35.4$\pm$3.0\\
&2017.12.31   & V 30$\times$3, R 30$\times$3, I 30$\times$3 &0.79&$-$18.0$\pm$8.5&37.5$\pm$3.4\\
&2018.01.03   & V 30$\times$3, R 30$\times$3, I 30$\times$3 &0.83&$-$17.6$\pm$1.4&29.8$\pm$3.1\\
&2018.01.10   & V 30$\times$3, R 30$\times$3, I 30$\times$3 &0.91&3.7$\pm$3.0&32.2$\pm$2.9\\
&2018.01.16   & V 30$\times$3, R 30$\times$3, I 30$\times$3 &0.99&23.8$\pm$3.7&29.4$\pm$3.4\\
&2018.01.20   & V 30$\times$3, R 30$\times$3, I 30$\times$3 &0.04&42.6$\pm$1.2&28.3$\pm$3.3\\
&2018.01.27   & V 30$\times$3, R 30$\times$3, I 30$\times$3 &0.13&62.5$\pm$3.2&28.4$\pm$3.2\\
&2018.01.28   & V 30$\times$3, R 30$\times$3, I 30$\times$3 &0.14&69.7$\pm$4.4&22.2$\pm$3.1\\
&2018.02.15   & V 30$\times$3, R 30$\times$3, I 30$\times$3 &0.37&72.7$\pm$2.1&25.8$\pm$3.0\\
&2018.03.04   & V 30$\times$3, R 30$\times$3, I 30$\times$3 &0.58&6.1$\pm$1.2&30.8$\pm$3.3\\
&2018.03.13   & V 30$\times$3, R 30$\times$3, I 30$\times$3 &0.70&$-$19.8$\pm$1.7&29.8$\pm$2.9\\
&2018.03.16   & V 30$\times$3, R 30$\times$3, I 30$\times$3 &0.74&$-$28.9$\pm$1.2&33.9$\pm$3.2\\
&2018.03.24   & V 30$\times$6, R 30$\times$6, I 30$\times$6 &0.84&$-$25.2$\pm$1.5&29.5$\pm$2.9\\
&2018.03.29   & V 30$\times$3, R 30$\times$3, I 30$\times$3 &0.90&1.6$\pm$1.4&35.4$\pm$3.6\\
&2018.04.07   & V 30$\times$3, R 30$\times$3, I 30$\times$3 &0.01&29.0$\pm$2.8&29.5$\pm$3.0\\
&2018.04.26   & V 30$\times$3, R 30$\times$3, I 30$\times$3 &0.26&77.4$\pm$3.7&22.6$\pm$3.0\\
\hline
\multirow{7}{0.5in}{Keck}& 2017.12.04   & 600 &0.44&52.8$\pm$1.4&26.0$\pm$1.5\\
& 2017.12.09  & 300 &0.50&32.7$\pm$1.4&26.5$\pm$1.4\\
& 2017.12.10  & 300 &0.51&28.2$\pm$1.4&31.7$\pm$1.5\\
& 2017.12.24  & 600 &0.69&$-$18.3$\pm$1.4&37.2$\pm$1.3\\
& 2017.12.29  & 600 &0.75&$-$22.9$\pm$1.4&37.0$\pm$1.4\\
& 2017.12.31  & 500 &0.78&$-$21.8$\pm$1.4&36.2$\pm$1.3\\
& 2018.01.06  & 600 &0.86&$-$13.5$\pm$1.4&34.5$\pm$1.3\\
\hline
\end{tabular}
\end{center}
See Methods section 'Discovery and follow up observations of LB-1', 'Radial velocity measurements' for details.\\
$^a$ The $RV_{\rm \alpha}$ corresponds to the 1/3 height method.
\end{table}

\begin{table}
\caption*{\bf Extended Data Table 2. H$\alpha$ measurement with different methods.}
\begin{center}
\renewcommand\arraystretch{0.8}
\begin{tabular}{cccccccc}
\hline
Width/method   &  $K_{\rm \alpha}$(km/s)  & \multicolumn{2}{c}{Uncertainty}  &$V_{\rm 0\alpha}$(km/s)   & \multicolumn{2}{c}{Uncertainty} \\
            &         & (90\%  ) &(99\%  ) &&(90\%  ) &(99\%  )\\
\hline
2/3  Height    &  3.9   &  0.8  &    1.2       &   29.1   &   0.5   &   0.9\\
1/2  Height    & 4.4 &   0.7    &    1.0        &   28.7  &  0.5   &  0.7\\
1/3  Height   &  6.4      &0.8   &    1.3 & 28.9   & 0.6  &    1.0\\
1/4  Height   &  5.8      &1.0   &    1.5 & 29.2   & 0.7  &    1.1\\
1/5  Height   &  6.7      &1.0   &    1.6 & 29.1   & 0.8  &    1.2\\
120km/s &  4.1  &  0.8 &      1.2 & 29.2 &    0.6&    0.9\\
140km/s &  4.8 &     0.8 &      1.3 &    29.0 &     0.6 &     0.9\\
170km/s &     5.5 &     0.8&      1.3&      29.3 &     0.6&     1.0\\
200km/s &    6.0 &     0.9 &   1.4&       29.4&     0.7 &      1.1\\
Bary center (no mask)   & 1.7&    0.9&      1.5&      29.5&     0.7&      1.1\\
\hline
\end{tabular}
\end{center}
See Methods section 'Radial-velocity measurements' for details.\\
\end{table}

\clearpage

\begin{table}
\label{par.tab}
\caption*{\bf Extended Data Table 3. Orbital parameters of LB-1.}
\begin{center}
\renewcommand\arraystretch{0.8}
\begin{tabular}{cccc}
\hline
Parameter   &  Value  & \multicolumn{2}{c}{Uncertainty}              \\
&         & (90\% confidence) &(99\% confidence) \\
(1)         & (2)  & (3) &  (4)\\
\hline
$e$             &   0.03  &  0.01  &  0.01\\
$K_{\rm B}$           &   52.8  &  0.7   &  1.0\\
$V_{\rm 0B}$           &   28.7  &  0.5   &  0.7\\
$K_{\rm \alpha}^a$ &   6.4   &   0.8  &  1.3\\
$V_{\rm 0\alpha}^a$ &   28.9  &   0.6  &   1.0\\
\hline
\end{tabular}
\end{center}
See Methods section 'Period and orbital parameters' for details.\\
$^a$ The $K_{\rm 0\alpha}$ and $V_{\rm 0\alpha}$ correspond to the 1/3 height method.
\end{table}


\clearpage
\newpage

\begin{figure}
\includegraphics[width=1\textwidth]{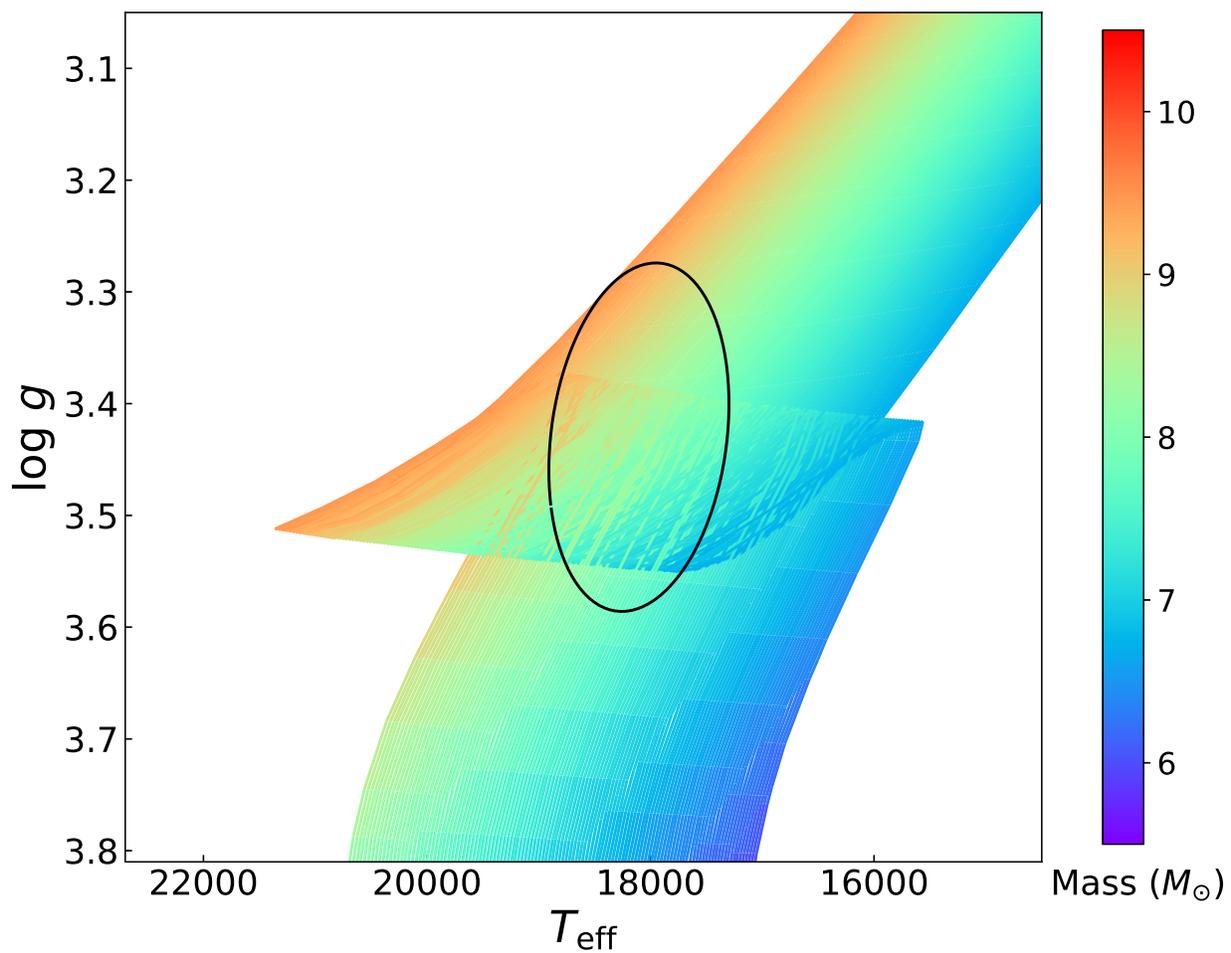}
\caption*{{\bf Extended Data Figure 1. Using isochrones from PARSEC models.} The grid of log$g$ and $T_{\rm eff}$ was constructed using the PARSEC isochrones. The black ellipse indicates 90\% uncertainty of the $T_{\rm eff}$ and log$g$ of the B star; all points inside it are considered as acceptable models for the B star.}
\label{mass.fig}
\end{figure}

\begin{figure}
\includegraphics[width=1\textwidth]{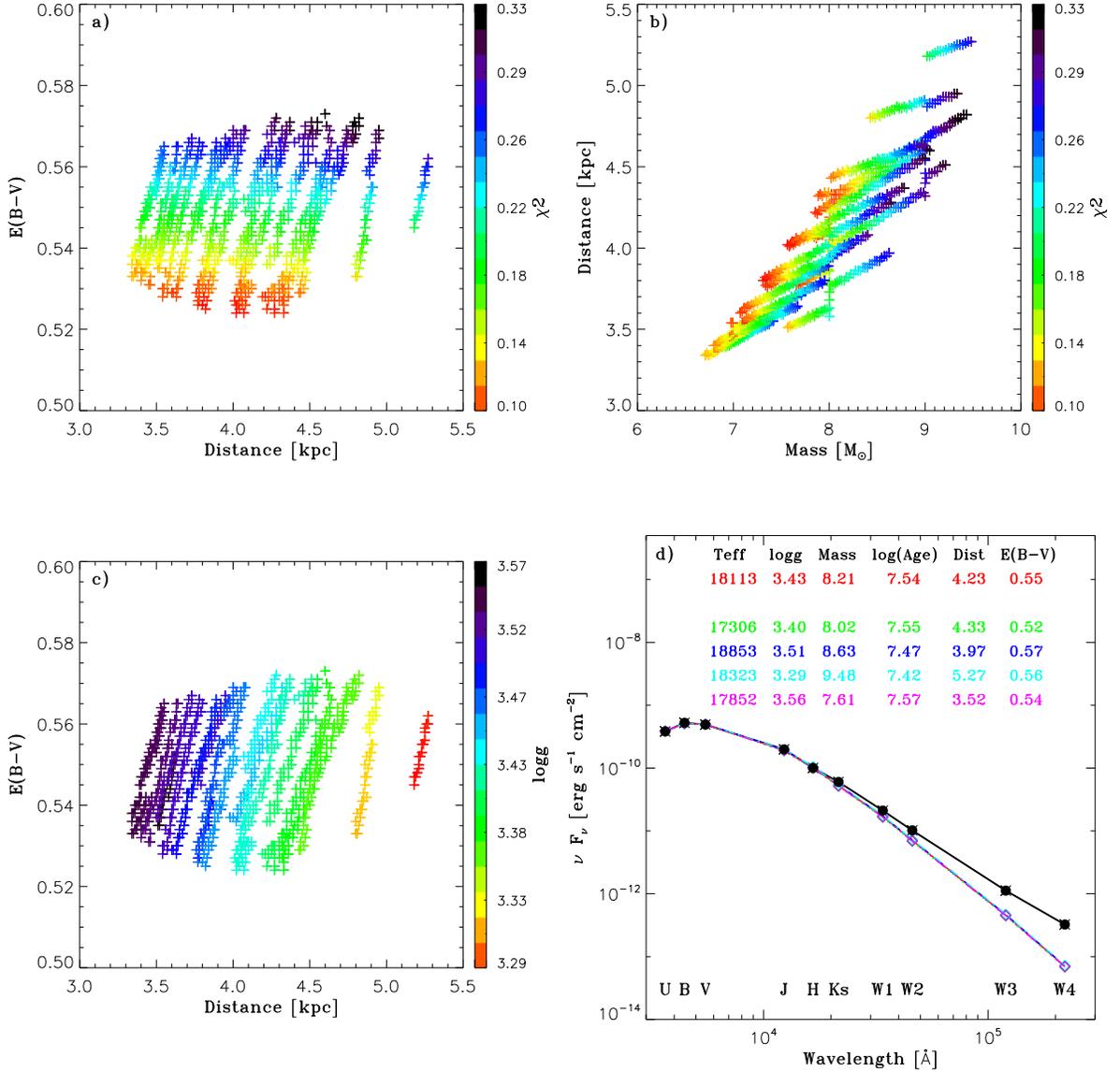}
\caption*{{\bf Extended Data Figure 2. SED fitting results for the B star.} {\bf a}, E(B-V) versus distance, both of which are from the SED fitting. The colorbar indicates the $\chi^2$. {\bf b}, Distance versus stellar mass, the latter being determined from the acceptable PARSEC models of the B star. The colour bar indicates $\chi^2$. {\bf c}, E(B-V) versus distance. The colour bar indicates log$g$. {\bf d}, Several examples of the SED fitting. The black squares are the data from the UCAC4, 2MASS, and AllWISE catalogues. See Methods for details.}
\label{sed.fig}
\end{figure}

\begin{figure}
\includegraphics[width=1\textwidth]{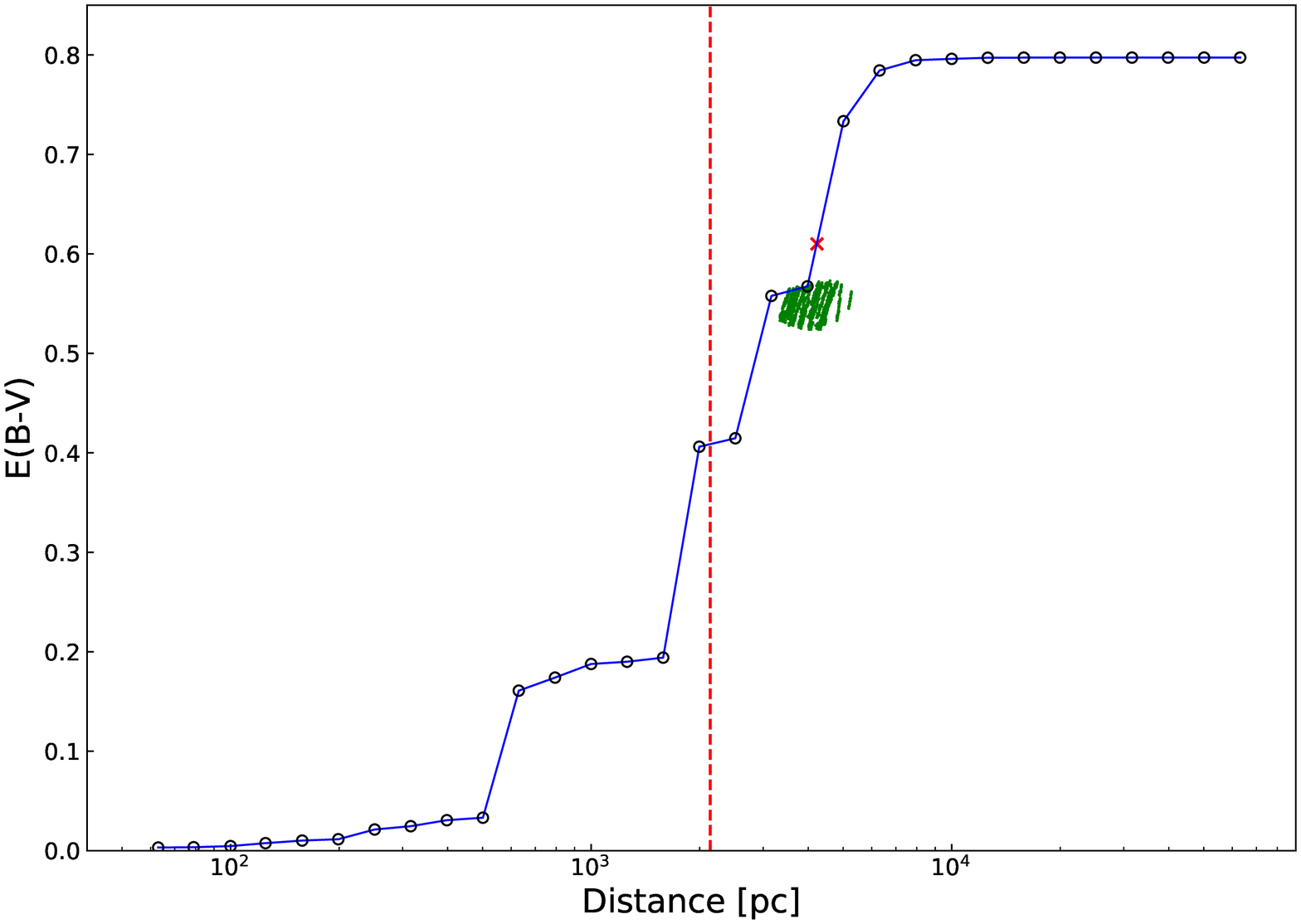}
\caption*{{\bf Extended Data Figure 3. Variation of $E(B-V)$ with distance in the direction of LB-1.} The black circles represent the extinction values corresponding to different distances from 3D dust map. The green points are the extinction and distances from SED fitting for each acceptable model of the B star. The red cross marks the extinction value from the 3D dust map at 4.23 kpc, while the red dashed line shows the Gaia distance of 2.14 kpc.}
\label{pan.fig}
\end{figure}

\begin{figure}
\includegraphics[width=1\textwidth]{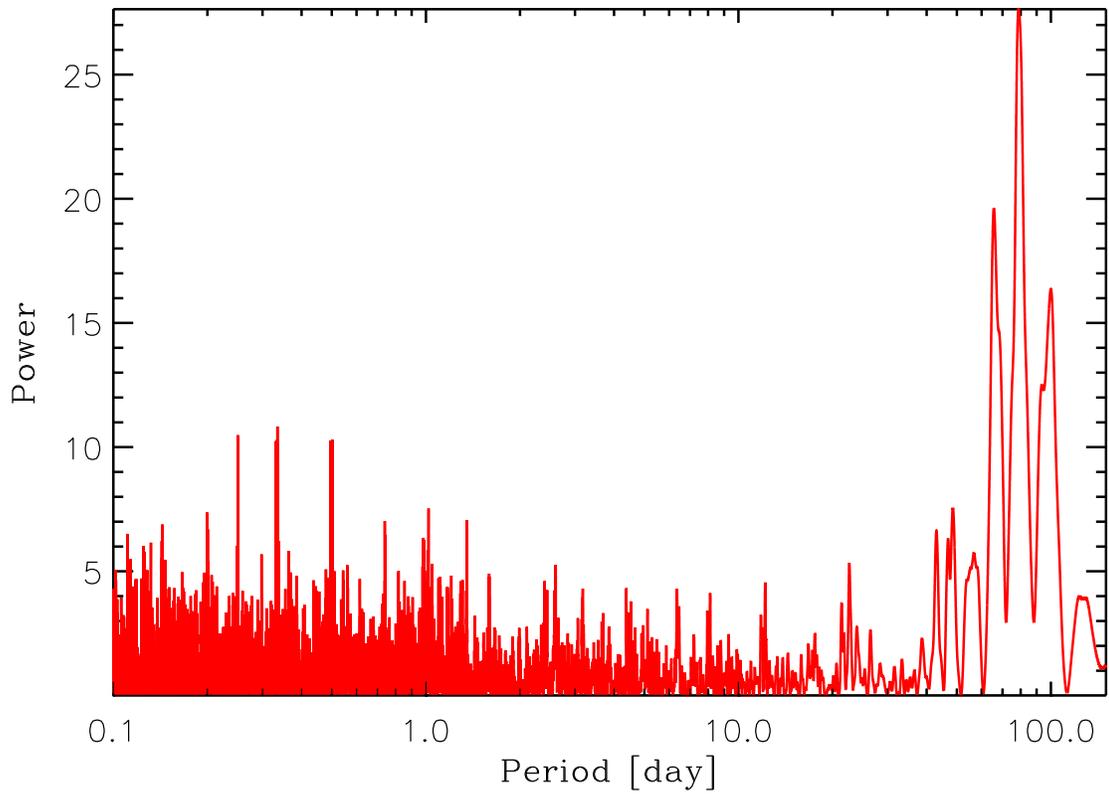}
\caption*{{\bf Extended Data Figure 4. Search for periodicities for LB-1 with the Lomb-Scargle method.} The
radial-velocity curve from LAMOST, GTC and Keck observations is being used here. The highest peak corresponds to the orbital period of $\approx$ 78.9\,day.}
\label{lomb.fig}
\end{figure}

\begin{figure}
\centering
\includegraphics[width=1\textwidth]{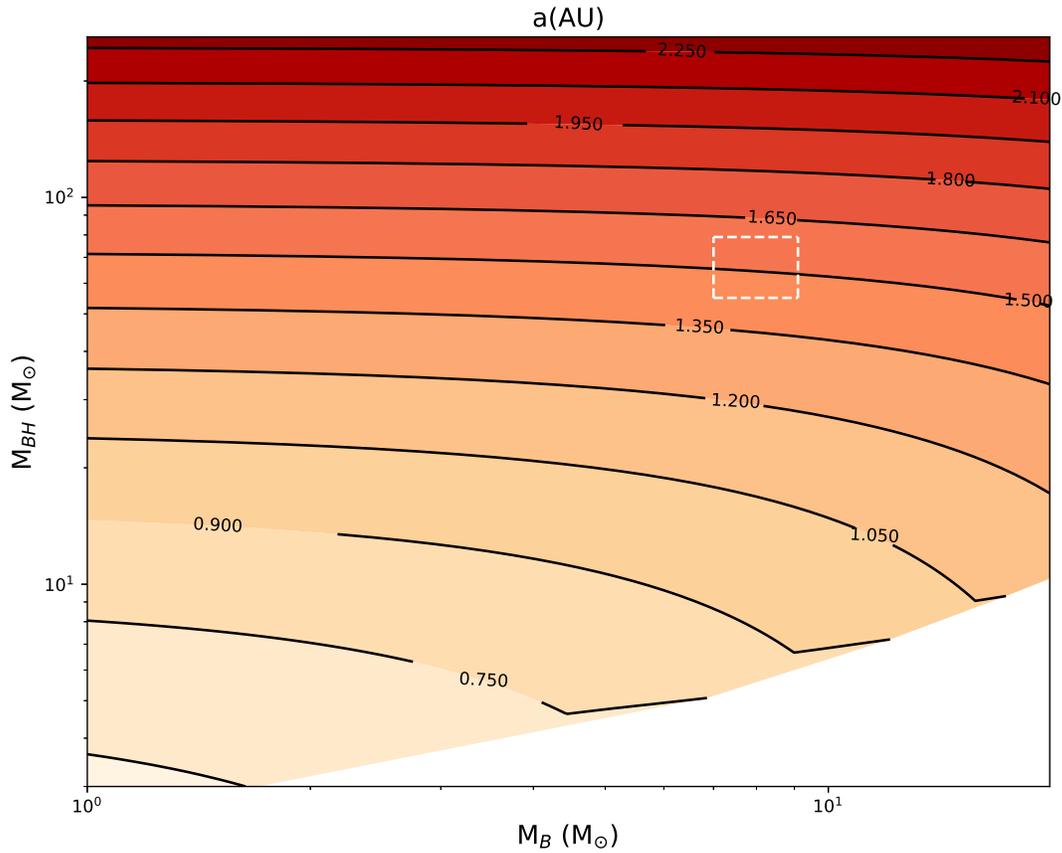}
\caption*{{\bf Extended Data Figure 5. Separation $a$ as a function of $M_{\rm B}$ and $M_{\rm BH}$.} Here $a$ is calculated from Kepler’s third law for each pair of $M_{\rm B}$ (B-star mass) and $M_{\rm BH}$ (black-hole mass). The contours and colours both represent the values of $a$. The white dashed lines in the contour plot outline a valid region of the separation of the binary system. It comes from the limitations for the $M_{\rm B}$ (7$\sim$9.1\,M$_{\odot}$) and the $M_{\rm BH}$ (55$\sim$79\,M$_{\odot}$).}
\label{yz3.fig}  
\end{figure}

\begin{figure}
\centering
\includegraphics[width=1\textwidth]{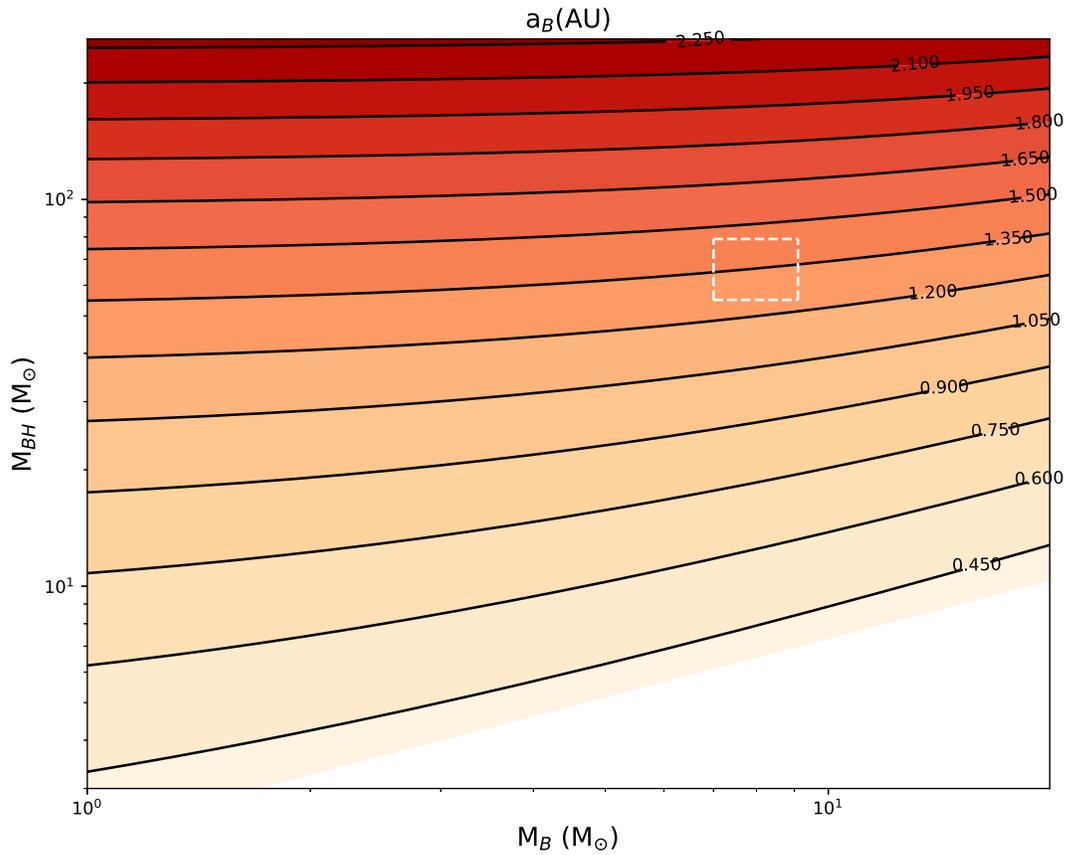}
\caption*{{\bf Extended Data Figure 6. Semi-major axis of the orbit of the B star $a_{\rm B}$ as a function of $M_{\rm B}$ and $M_{\rm BH}$.} Here $a_B$ is calculated from Kepler’s third law for each pair of $M_{\rm B}$ (B-star mass) and $M_{\rm BH}$ (black-hole mass). The contours and colours both represent the values of $a_B$. The white dashed lines in the contour plot outline a valid region for the semi-major axis of the B star. It comes from the limitations for the $M_{\rm B}$ (7$\sim$9.1\,M$_{\odot}$) and  the $M_{\rm BH}$ (55$\sim$79\,M$_{\odot}$).}
\label{yz4.fig}  
\end{figure}

\begin{figure}   
\includegraphics[width=1\textwidth]{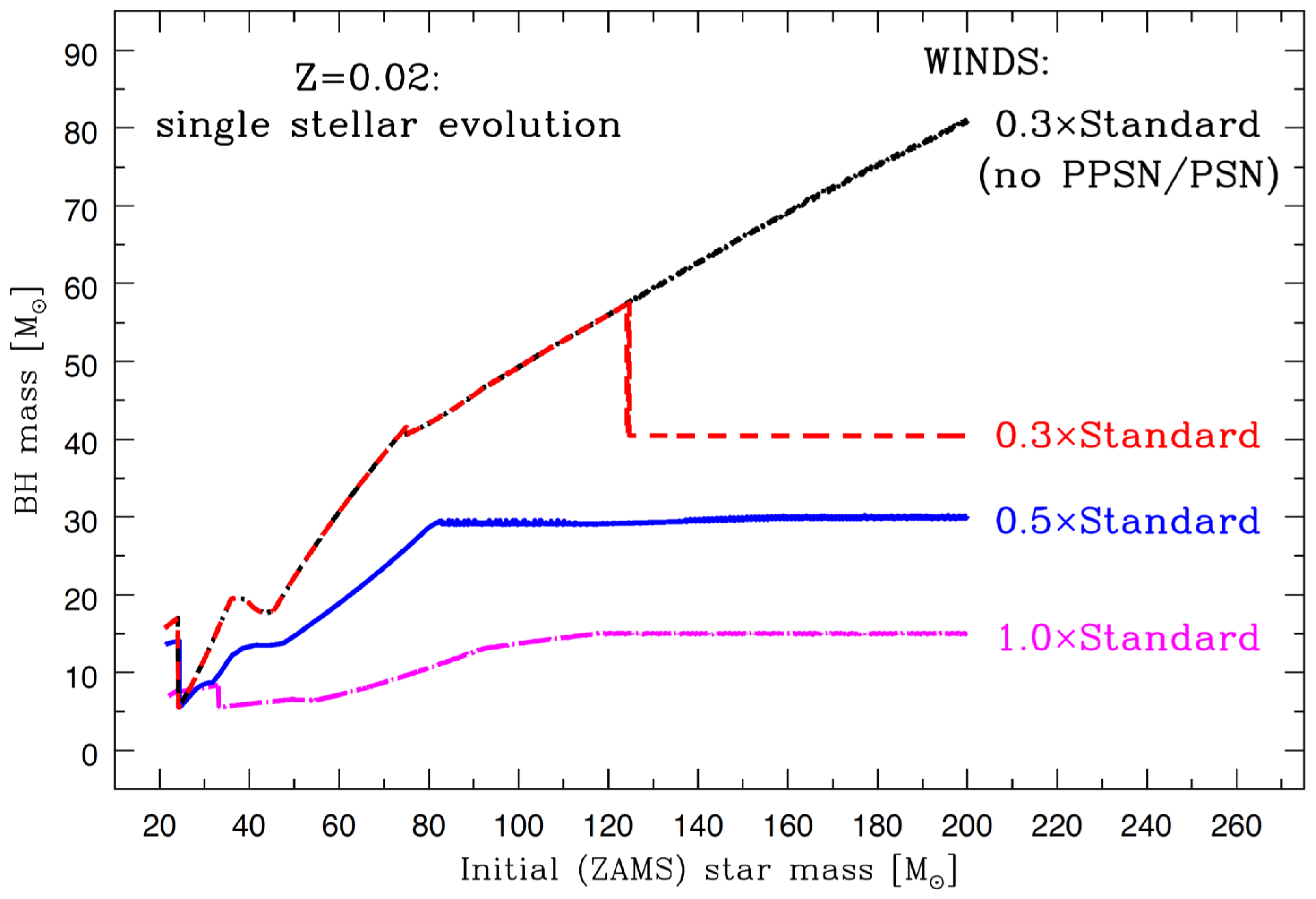}  
\caption*{{\bf Extended Data Figure 7. Black hole mass versus initial mass in the zero age main sequence (ZAMS) for single stars.} For standard wind mass loss prescriptions only low-mass black holes are predicted: $M_{\rm BH}<15$\,M$_{\odot}$. However, for reduced wind mass loss much heavier black holes are formed: $M_{\rm BH}=30$\,M$_{\odot}$ for winds reduced to $50\%$, and $M_{\rm BH}=60$\,M$_{\odot}$ for winds reduced to $30\%$ of the standard values. Note that to reach $M_{\rm BH}=80$\,M$_{\odot}$ it is needed to switch off pair-instability pulsation supernovae (PPSN) or pair-instability supernovae (PSN), which severely limit black hole masses.}
\label{kb1.fig}
\end{figure}

\end{document}